\documentclass{aa}

\usepackage[utf8]{inputenc}
\usepackage[varg]{txfonts}
\usepackage{amssymb}
\usepackage{epsfig}
\usepackage{graphics}
\usepackage{amsmath}
\usepackage{color}
\usepackage{natbib}
\usepackage{hyperref}
\usepackage{gensymb}

\usepackage{aas_macros}

\usepackage{cleveref}
\usepackage{soul}

\usepackage{bm}
\usepackage{mathtools}
\usepackage{graphicx}
\usepackage{lipsum}
\usepackage{physics}
\usepackage{ulem}

\definecolor{yligreen}{rgb}{0.5,0.5,0.0}

\newcommand{\cbrt}[1]{\sqrt[\scriptstyle 3]{#1}}
\newcommand{\rmd}{{\rm d}}
\newcommand{\rs}{{R}_{\rm S}}
\newcommand{\risco}{{r}_{\rm{ISCO}}}
\newcommand{\unit}[1]{\mbox{\boldmath $\hat{#1}$}}

\bibpunct{(}{)}{;}{a}{}{,} 

\DeclareUnicodeCharacter{00A0}{ }

\makeatletter
\def\fvec#1{\underline{\sbox\tw@{$#1$}\dp\tw@\z@\box\tw@}}
\makeatother

\begin{document}
\title{\textsc{artpol}: 
Analytical ray-tracing method for spectro-polarimetric properties of accretion disks around Kerr black holes}

\titlerunning{Polarization properties of Kerr accretion disks}

\author{Vladislav~Loktev\inst{1}
\and  Alexandra~Veledina\inst{1,2}
\and  Juri~Poutanen\inst{1}
\and Joonas N\"attil\"a\inst{3,4}
\and Valery F. Suleimanov\inst{5}
}

\institute{Tuorla Observatory, Department of Physics and Astronomy, FI-20014 University of Turku, Finland \email{vladislav.loktev@utu.fi}
\and Nordita, KTH Royal Institute of Technology and Stockholm University, Hannes Alfv\'{e}ns v\"{a}g 12, SE-10691 Stockholm, Sweden
\and Physics Department and Columbia Astrophysics Laboratory, Columbia University, 538 West 120th Street New York, NY 10027
\and Center for Computational Astrophysics, Flatiron Institute, 162 Fifth Avenue, New York, NY 10010, USA
\and Institut f\"ur Astronomie und Astrophysik, Universit\"at T\"ubingen, Sand 1, D-72076 T\"ubingen, Germany
}


\abstract{Spectro-polarimetric signatures of accretion disks in X-ray binaries and active galactic nuclei contain information about the masses and spins of their central black holes, as well as the geometry of matter close to the compact objects.
This information can be extracted using the means of X-ray polarimetry.
In this work, we present a fast analytical ray-tracing technique for polarized light \textsc{artpol} that helps obtain the spinning black hole parameters from the observed properties.
This technique can replace the otherwise time-consuming numerical ray-tracing calculations.
We show that \textsc{artpol} proves accurate for Kerr black holes with dimensionless spin parameter $a\leq0.94$ while being over four orders of magnitude faster than direct ray-tracing calculations.
This approach opens broad prospects for directly fitting the spectro-polarimetric data from the \textit{Imaging X-ray Polarimetry Explorer}.}


\maketitle

\section{Introduction}\label{sec:intro}

The spin of a black hole (BH) is a fundamental parameter that controls the behavior of the inflowing matter, the accretion disk, and the properties of the outflowing material, the relativistic jets.
The magnitude of spin determines the curvature of space-time close to the BH, determines the energy dissipation profile, and affects the spectral energy distribution of the observed emission.
The spin values for both Galactic BHs in X-ray binaries and their supermassive counterparts have been probed using the distinct imprints in spectral and timing properties via the continuum-fitting method and the iron line reflection/reverberation method \citep{Miller2009,McClintock14,Uttley2014}.
The methods are based on obtaining the radius of the innermost stable circular orbit (ISCO), which, in turn, is related to the BH spin.
For X-ray binaries, an additional constraint on the parameters comes from the relativistic precession model \citep{Motta2014,Motta2022}, which links the mass, spin, and radius with the characteristic frequencies found in the X-ray light curves.
The statistical distribution of spins probed by these methods, as well as spin values obtained by different methods for the same source, do not match \citep{Draghis2023}.
This calls for an alternative method to verify spin determination measurements.

Polarization of radiation escaping from the BH vicinity can be used as a fine tool to determine the curvature of space-time.
In this context, X-ray polarimetric signatures of accretion disks have long been anticipated to carry important information about the BH spin, and the launch of \textit{Imaging X-ray Polarimetry Explorer} \citep[\textit{IXPE},][]{Weisskopf2022} has opened these new exciting possibilities.
At the same time, the observed polarimetric signatures of BH X-ray binaries have proved many previously proposed models to fail in describing the data \citep{Krawczynski22,RodriguezCavero2023,Ratheesh2023}, requiring to alter basic assumptions on the geometry and radiative mechanisms producing local spectra, which are often impossible to alter in the data-fitting models.
A fast tool that relates the local spectra with the observed spectro-polarimetric signatures is required.

The accretion disk polarization is produced by multiple scatterings in the upper layers of its atmosphere.
The first predictions of the disk polarization \citep{Rees1975} have been made using the results of calculations in the case of pure electron-scattering, semi-infinite plane-parallel atmospheres \citep{Chandrasekhar1947,sob49,Cha60,Sob63}, and were limited to Newtonian approximation.
The polarization degree (PD) may be altered due to the presence of absorption effects in the atmosphere, which were considered using Monte-Carlo \citep{LightmanShapiro1975} and analytical \citep{Losob79,Losob81} means.
Further, the effects of general and special relativity introduce important modifications to both PD and polarization angle \citep[PA;][]{ConnorsStark1977,StarkConnors1977,PineaultRoeder1977kerr_analyt,PineaultRoeder1977kerr_num}.
Aberration and light deflection lead to a rotation of the PA and alter the viewing angle of different parts of the accretion disk, affecting the PD. 
Additional rotation of PA along the photon trajectory is expected for the case of spinning BHs, described by the \citet{Kerr1963} metric, thanks to the frame-dragging effects.
Therefore, the observed spectral dependence of polarimetric signatures can act as an independent probe of the BH spin \citep{ConnorsPiranStark1980,Dovciak2008}.

Precise computations of the effects of general and special relativity on the polarization properties can be done using the parallel transport of the polarization vector along null geodesics, which often involves computations of the \citet{WalkerPenrose1970} constant of motion \citep{ConnorsPiranStark1980,Dovciak2008,Ingram15}.
The geodesics, in turn, are computed using the ray-tracing techniques \citep[e.g.,][]{Dexter2016,MONK,GRay,PMN18}.
It is also essential to keep track of the convergence of the computed flux, PD and PA down to the characteristic scale of the observed errors on these quantities--this often means that several simulations with increased resolution must be performed for each parameter set.
An additional source of computational errors comes from the assumptions of a small outer radius of the disk, which is enforced by the high computational costs.
Typically, the value $\lesssim100R_{\rm S}$ (where $R_{\rm S}$ is the Schwarzschild radius) is considered, which allows for the (highly polarized) secondary images of the disk to be visible in the region surrounding the outer radius of the simulated disk -- in reality, a much higher extend of the disk, $\gtrsim10^{5}R_{\rm S}$, completely covers those from the line of sight.

The ray-tracing technique is too computationally expensive for a direct data fitting.
Instead, pre-computed geodesics have been used to accelerate calculations \citep[e.g.,][]{2005ApJS..157..335L, Krawczynski2012,kerrC}.
Alternatively, analytical and semi-analytical approaches can be applied to solve geodesic equations \citep{Dexter2009,ynogkm,AART}. 
Their applicability is, however, limited, e.g., the semi-analytical expressions for geodesics can only be used for the equatorial plane of the BH.
Finally, in the Schwarzschild metric, the calculation of geodesics may be omitted, and the photon trajectory can be treated using an approximation to the light bending relation \citep{B02,Poutanen2020bending}.
The latter approach gave reliable results for the low-energy synchrotron emission observed from the supermassive BH M87*, with an almost face-on disk \citep{Narayan2021}.

Physical understanding of the modifications to polarization signatures caused by the curved space-time and fast motions of matter is difficult to achieve if we use the implicit Walker-Penrose constant.
For this, one can use the explicit analytical expression for the rotation of the polarization plane along the photon path, which we call the analytical ray-tracing technique for polarized light (\textsc{artpol} hereafter).
This approach has been used to extract polarimetric properties of spinning spherical \citep{poutanen20} and oblate \citep{Loktev20} neutron stars, as well as accretion disks around Schwarzschild BHs \citep{Loktev2022}.
The method was first proposed in \citet{Pineault1977pol_schw} and applied, in the context of accretion disks, to the PA rotation caused by, separately, general and special relativity, while the expression for their combined effects was first derived in \citet{Loktev2022}.

In this work, we apply \textsc{artpol} to the accretion disks around spinning BH. 
In Sect.~\ref{sec:methods}, we describe the formalism that can be used in spectro-polarimetric modeling and imaging of accretion disks around Kerr BHs.
In Sect.~\ref{sec:results}, we compare the results of the \textsc{artpol} technique to those obtained using explicit ray-tracing calculations.
We show that the PD and PA computed using this method remain accurate to the level of current observational  (\textit{IXPE}) uncertainties for BH spin parameters up to $a=0.94$.
We summarize our findings and discuss a broad range of applications of the technique in Sect.~\ref{sec:summary}.

\section{Methods}\label{sec:methods}

\subsection{Local model of the disk emission}\label{sec:local_disk}

We consider a BH with mass $M$ and dimensionless spin $a=Jc/GM^2$, where $J$ is the angular momentum. 
For the sake of simplicity, in this paper, we consider a standard equatorial geometrically thin accretion disk \citep{NT73}. 
The BH is situated at the origin with the spin directed along the $z$-axis and is orthogonal to the disk plane.
We assume that the state of the fluid only depends on the Boyer-Lindquist dimensionless radial coordinate $r=R/\rs$, which expresses the distance from the central object $R$ in units of the BH Schwarzschild radius $\rs=2GM/c^2$.
At a given radius $r$, the matter moves, relative to the locally non-rotating observer, with Keplerian velocity, which we express in the units of the speed of light $c$ as \citep{KFM08}: 
\begin{equation}\label{eq:kerrbeta}
    \beta = \frac{\cal F}{\cal B \sqrt{\cal D} } \sqrt{\frac{1}{2r}},
\end{equation}
where
\begin{align}
\cal B  &= 1 + \frac{a}{\sqrt{8r^3}}, \\
\cal D &=  1 -  \frac{1}{r} + \frac{a^2}{4r^2},\\
{\cal F} &= 
1 - \frac{a}{\sqrt{2r^3}}  + \frac{a^2}{4r^3} . 
\end{align}
The disk is considered to be optically thick.
The energy flux from a surface element is described by the effective temperature, which has the radial profile in the form
\begin{equation}\label{eq:tempprofile}
    T_{\rm eff}^4(r) = \frac{3GM\dot{M}}{8\pi \sigma_{\rm SB} R^3  } f(r,a) = T_*^4 \frac{f(r,a)}{r^3},
\end{equation}
where
\begin{equation}\label{eq:tempstar}
    T_*^4  = \frac{3GM\dot{M}}{8\pi \sigma_{\rm SB} \rs^3} ,
\end{equation} 
combines the BH mass and the accretion rate $\dot{M}$, and $f(r,a)$ is the factor accounting for the relativistic and boundary condition corrections (\citealt{PT1974}; Eq.~(3.191) in \citealt{KFM08}).
In the Newtonian case, assuming the inner disk edge at $r=3$, the correction factor is
\begin{equation}\label{eq:corr_Newton}
     f(r) = 1-\sqrt{\frac{3}{r}}.
 \end{equation} 
The disk extends from the ISCO up to an outer edge at $r_{\rm{out}}=3000$, which was chosen to reproduce the correct spectra down to the photon energy $E\sim 0.01 kT_*$. 
The radius of the ISCO depends on the spin as
\begin{equation}\label{eq:risco}
\risco = \frac{1}{2}\left( 3 + Z_2 \pm \sqrt{(3-Z_1)(3+Z_1+2Z_2)} \right) , 
\end{equation}
where the minus sign corresponds to a corotation of the disk and the BH and the plus sign is for the case of the retrograde rotation, and 
\begin{align}
Z_1 &= 1 + \cbrt{1-a^2} \left( \cbrt{1+a} + \cbrt{1-a} \right) , \\
Z_2 &= \sqrt{3a^2 + Z_1^2} .
\end{align}

The local spectrum of the disk is assumed to be a diluted blackbody with the color correction $f_{\rm col} = 1.7$ \citep{ST95} with the flux being 
\begin{equation}
    F_{E'} = \frac{\pi}{f_{\rm col}^4} B_{E'}(f_{\rm col}T_{\rm eff}) , 
\end{equation}
where ${E'}$ is photon energy measured in the frame comoving with the matter of the disk (where all quantities are denoted with primes) and $B_{E'}$ is the Planck function. 
The angular distribution of the specific intensity and the polarization is assumed to correspond to the case of electron-scattering-dominated semi-infinite atmosphere \citep{Chandrasekhar1947,Cha60,sob49,Sob63}.  
The three-component Stokes vector fully describes a linearly polarized radiation field as a function of photon energy $E'$ and zenith angle
$\zeta'$ (measured in the comoving frame): 
\begin{equation} \label{eq:primepolvec}
\bm{I}'_{E'}(\zeta')= 
\begin{bmatrix} I'_{E'} \\ Q'_{E'} \\  U'_{E'} \end{bmatrix} = \frac{1}{{\pi}}
F_{E'}   a_{\rm es}(\zeta')
\begin{bmatrix} 1\\ p_{\rm es}(\zeta') \cos2\chi_{0} \\ p_{\rm es}(\zeta') \sin2\chi_{0} \end{bmatrix} ,
\end{equation}
where the angular distribution can be approximated as \citep{NP18}
\begin{equation}
\label{eq:Chand_angle}
a_{\rm es}(\zeta') =  \frac{60}{143} (1+2.3\cos{\zeta'}-0.3\cos^2{\zeta'})    
\end{equation}
and the PD \citep{VP04}
\begin{equation}
\label{eq:Chand_PD}
p_{\rm es}(\zeta') = 0.1171\frac{1-\cos{\zeta'}}{1+3.582\cos{\zeta'}} . 
\end{equation}
The Stokes parameter $U$ is zero in the comoving frame because of the azimuthal symmetry. 
Furthermore, in our case, the escaping radiation is polarized perpendicular to the meridional plane formed by the normal to the local surface and the photon momentum, resulting in the local PA of $\chi_0=\pi/2$ (Stokes $Q<0$).

Once the model for the disk structure (i.e., radial dependence of the velocity profile) and the energy and angular dependence of the local Stokes vector are established, we need to consider how the polarized radiation is modified toward the observer in the curved space-time. 
One way to account for that is to compute a set of geodesics from the vicinity of the BH to the observer (we focus on this in more detail in Sect.~\ref{ssec:rt}).
The other possibility is to use a faster approach that exploits analytical approximations, which we describe in Sect.~\ref{ssec:analyt}.

\begin{figure} 
\centering
\includegraphics[width=0.95\columnwidth]{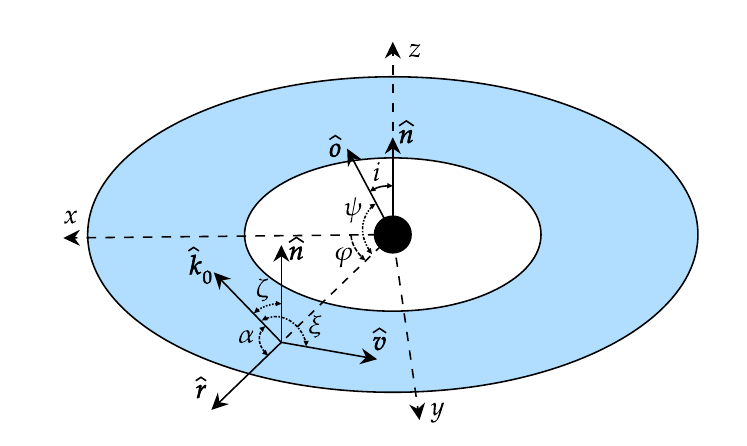}
\caption{Geometry of a flat accretion disk ring.
An element of the flat equatorial disk, defined by the unit radius-vector $\unit{r}$ at azimuthal angle  $\varphi$, moves along the direction $\unit{v}$. 
The coordinate system is based on the normal of the disk $\unit n$ and the observer vector $\unit o$.
A photon is emitted from the disk element along $\unit{k}_{0}$.
The important angles used in this work are also shown.
} 
\label{fig:geom}
\end{figure}

\subsection{Analytical ray-tracing in Schwarzschild metric} 
\label{ssec:analyt}

We previously tested the analytical method of calculating the rotation of the PA for the case of the planar light trajectories \citep{Loktev2022}.
This assumption is fulfilled in the Schwarzschild metric but does not hold for the Kerr metric.
Nevertheless, we can use this approach as the first-order approximation for the case of a spinning BH.

The considered geometry is shown in Fig.~\ref{fig:geom}. 
The disk is assumed to be flat and located in the equatorial plane of the spinning BH.
An element of the accretion disk surface, located at the tip of the radius vector $\bm{r}$, is described in Boyer-Lindquist coordinates by an azimuth angle $\varphi$ and the length $r$.   
We utilize a 3-dimensional Cartesian coordinate system to describe the polarization frame's rotation. 
The normal to the disk $\unit{n}$ is aligned with the $z$ axis and the direction to the observer $\unit{o}$ is in the $x$-$z$ plane making an inclination $i$ to the normal: 
\begin{eqnarray}\label{eq:geom_vectors}
\unit{n} & = & (0,0,1),  \\
\unit{o} & = & (\sin i,0,\cos i). 
\end{eqnarray}
In these coordinates, the unit vector of the disk element 
\begin{equation}\label{eq:diskele_vector}
\unit{r}  =  (\cos\varphi,\sin\varphi,0)
\end{equation}
makes an angle $\psi$ with the the observer direction 
\begin{equation}
\cos\psi  = \unit{r}\cdot \unit{o} = \sin i \cos \phi .
\end{equation} 
Close to the disk surface, photon escapes along a unit vector
\begin{equation}\label{eq:k0}
\unit{k}_0=[ \sin\alpha\ \unit{o} +\sin(\psi-\alpha)\ \unit{r}]/\sin\psi ,  
\end{equation} 
where $\alpha$ is the angle between the radius vector and the observer direction, 
\begin{equation}
\cos\alpha  = \unit{r}\cdot \unit{k}_0 ,
\end{equation}  
and is related to $\psi$ through the light bending formula (see below).
The photon momentum makes an angle $\zeta$ with the normal 
\begin{equation}
\cos\zeta  = \unit{n}\cdot \unit{k}_0 = 
 \frac{\sin\alpha}{\sin\psi}\cos i . 
\end{equation} 
Here we assume that the fluid velocity is aligned with the azimuthal unit vector  
\begin{equation}\label{eq:varphi_vector}
\unit{\varphi} = (-\sin\varphi,\cos\varphi,0) = \unit{v}.
\end{equation}
The matter moves with Keplerian speed given by Eq.~\eqref{eq:kerrbeta}.

The unit vector of photon momentum in the laboratory frame $\unit{k}_0$ and the one in the comoving frame $\unit{k}^{\prime}_0$ are related by the Lorentz transformation:
\begin{equation} \label{eq:k0Lorentz}
\unit{k}^{\prime}_0 = \delta \left[ \unit{k}_0 - \gamma\beta \unit{v} + (\gamma-1) \unit{v} (\unit{v}\cdot \unit{k}_0 ) \right],
\end{equation}
where $\gamma = 1/\sqrt{1-\beta^2}$ is the Lorentz factor and 
\begin{equation} \label{eq:doppler}
\delta = \frac{1}{\gamma(1 - \beta\cos\xi)} =  \gamma(1 +\beta\cos\xi')
\end{equation}
 is the Doppler factor and  
\begin{eqnarray}
 \cos\xi & = & \unit{v} \cdot \unit{k}_0  = 
 - \frac{\sin\alpha}{\sin\psi}\sin i \sin\phi, \\
\cos\xi' & = & \unit{v} \cdot \unit{k}^{\prime}_0 =  \frac{\cos\xi-\beta}{1-\beta\cos\xi} .
\end{eqnarray}
We then also get 
\begin{equation} \label{eq:coszetapr}
\cos\zeta' = \unit{n}\cdot \unit{k}^{\prime}_0 = \delta \cos\zeta. 
\end{equation}

The geodesics are not computed explicitly -- we only need the relation between the angles $\alpha$ and $\psi$, for which we use the following analytical approximate formula \citep{Poutanen2020bending}: 
\begin{equation}\label{eq:bendapprox}
\cos\alpha = 1- (1-u)y\left\{ 1+\frac{u^2y^2}{112}- \frac{{\rm e}}{100}uy \left[ \ln\left(1-\frac{y}{2}\right) +\frac{y}{2} \right] \right\},
\end{equation}
where $y = 1-\cos\psi$ and $u = 1/r$. 
The formula is not applicable for $r<1$ (within the Schwarzschild radius), hence for the cases with $\risco<1$, i.e., $a \gtrsim 0.943$, \textsc{artpol} cannot be used.

Under the Schwarzschild metric assumption, the geodesics are flat; therefore, the parallel transport of the Lorentz frame along the geodesics is unnecessary. 
The rotation of the polarization basis is a sum of several simple rotations due to the gravitational light bending (the general relativity effect) and Lorentz aberration (the special relativity effect), 
\begin{equation}\label{eq:chitot}
 \chi^\text{tot} = \chi^{\text{GR}} + \chi^{\text{SR}} . 
\end{equation} 
Analytical expressions for those have been derived in \citet{Loktev2022}:  
\begin{equation}\label{eq:tanGR}
\tan\chi^{\text{GR}} = \frac{  \cos i \sin \varphi   }{
     \cos\varphi + \tilde{a}\sin i },
\end{equation}
where 
$\tilde{a} = (1-\cos\alpha  \cos\psi)/(\cos\alpha - \cos\psi)$,
and
\begin{equation}\label{eq:tanSRGR}
 \tan\chi^{\text{SR}}
 =   -\beta\ \frac{\cos \alpha \cos\zeta}{\sin^2\zeta-\beta \cos\xi} .
\end{equation}
Expressions for $\chi^{\text{GR}}$ and $\chi^{\text{SR}}$ in vector form can be found in \citet{Loktev2022}.

The PA is measured in the main polarization basis, formed by the disk normal and the observer vector:
\begin{equation}\label{eq:polbas_on}
\unit{e}_1 = \frac{\unit{n}-\cos{i}\  \unit{o}}{\sin{i}},\qquad
\unit{e}_2 = \frac{\unit{o} \times \unit{n}}{\sin{i}} .
\end{equation}
The total PA in this basis is given by the sum of the relativistic rotations and the intrinsic PA $\chi_0$,
\begin{equation}\label{eq:chitot}
    \chi =\chi_0 + \chi^\text{tot} 
    = \chi_0 + \chi^{\text{SR}} + \chi^{\text{GR}}.
\end{equation}
The PA $\chi_0$ in the comoving frame is described in the polarization basis that is formed by the local normal vector $\unit{n}$ and the photon momentum vector in that frame $\unit{k}^{\prime}_0$:
\begin{equation}\label{eq:polbas_k0n}
\unit{e}_1' = \frac{\unit{n}-\cos{\zeta'}\  \unit{k}_0'}{\sin{\zeta'}},\qquad
\unit{e}_2' = \frac{\unit{k}_0' \times \unit{n}}{\sin{\zeta'}} .
\end{equation}
In our case of electron scattering atmosphere, the intrinsic PA $\chi_0=\pi/2$.

The area of an element of the disk surface is expressed as 
\begin{equation}
\rmd S  = \frac{\rs^2 r \, \rmd r \, \rmd \varphi}{ \sqrt{{\cal D}}} .
\end{equation}
It occupies the solid angle $\rmd \Omega$ on the sky and can be computed in Schwarzschild approximation as 
\begin{equation}\label{eq:solidElement}
\rmd \Omega =\frac{ \rmd S \cos\zeta}{D^2}   {\cal L} = \frac{\rs^2}{D^2} \frac{ r \, \rmd r \, \rmd \varphi }{\sqrt{\cal D}} {\cal L} \cos\zeta  ,
\end{equation}
where $D$ is the distance to the source and the lensing factor ${\cal L}$ is defined for Schwarzschild space-time as \citep{B02}
\begin{equation}
  {\cal L}  =  \frac{1}{1-u} \frac{\rmd \cos\alpha}{\rmd \cos\psi}.
\end{equation}
It can be computed analytically  \citep{Poutanen2020bending}
\begin{equation}\label{eq:lensing_app}
{\cal L} =  1 + \frac{3u^2y^2}{112}  - \frac{{\rm e}}{100} u y \left[ 2\, \ln\left(1-\frac{y}{2}\right) + y\frac{1-3y/4}{1-y/2}\right]
\end{equation}
following Eq.~\eqref{eq:bendapprox}.

Next, the intensity in the comoving frame ${I}'_{E'}$ is related to the observed one as  
\begin{equation}
\label{eq:inten_lorentz}
    \bm{I}_{E} =g^3 \bm{I}'_{E'},
\end{equation}
where $g$ is the redshift factor, which can be expressed in the case of a disk in the equatorial plane of the BH as
(see Eq.~C13 in \citealt{2005ApJS..157..335L})
\begin{equation}\label{eq:redshift}
 g = E/E'= \gamma \left[{\cal X} + {\cal Y}\beta + ({\cal X}\beta+{\cal Y})\cos\xi'\right],
\end{equation}
where
\begin{align}
{\cal X} &= \sqrt{\cal D /\cal A \phantom{.}},\\
{\cal Y} &= a/\sqrt{4r^4\cal A \phantom{.}}, \\
\cal A &=  1+(r+1)\frac{a^2}{4r^3}.  
\end{align}
In the case of $a=0$, the redshift factor is reduced to $g = \delta\sqrt{1-1/r}$.

The total disk flux is obtained by integration over the surface: 
\begin{eqnarray}
\label{eq:polflux}
 \bm{F}_E & \equiv &  \begin{bmatrix} F_I(E)\\ F_Q(E) \\ F_U(E) \end{bmatrix} = 
\int \rmd \Omega \begin{bmatrix} I_{E}\\ Q_{E} \\ U_{E} \end{bmatrix} \nonumber \\
& = &   \frac{\rs^2}{D^2} 
\int\limits_{\risco}^{r_\textrm{out}}
  \frac{r\, \rmd r}{\sqrt{\cal D}}  \int_0^{2\pi}
     \rmd \varphi  \, g^{3} {\cal L} \cos\zeta\  {\bf M}(r,\varphi) \bm{I}'_{E'}(\zeta') ,
\end{eqnarray}
where ${\bf M}$ is the rotation matrix defined at every surface element of the disk, $\rmd r \, \rmd \varphi$, as following 
\begin{equation}\label{eq:pol-rot-matr}
    {\bf M}(r,\varphi) = \begin{bmatrix}
        1 & 0 & 0 \\
        0 & \cos2\chi^{\rm tot} & -\sin2\chi^{\rm tot} \\
        0 & \sin2\chi^{\rm tot} & \cos2\chi^{\rm tot}
    \end{bmatrix} .
\end{equation}

The PD and PA are then defined from the Stokes parameters of the total flux.
The PD is obtained as
\begin{equation}
p(E)=\frac{\sqrt{F ^2_{Q}(E)+F^2_{U}(E)}}{F_{I}(E)},
\end{equation}
while the observed PA can be computed from either 
\begin{equation}
\tan{2\chi(E)}=\frac{F_{U}(E)}{F_{Q}(E)}
\end{equation}
or
\begin{equation}
\tan{\chi(E)}=\frac{p(E) F_{I}(E) - F_{Q}(E)}{F_{U}(E)} .
\end{equation}

Following  \citet{Loktev2022}, we can use dimensionless energy $x = E/kT_*$ (and $x' = x/g$) and scale the luminosity to $\sigma_{\rm SB}T_*^4\rs^2$ to get the dimensionless luminosity $l_x$ in the following form
\begin{eqnarray}
 \label{eq:lum_scaled}
x \bm{l}_x & = &  
\frac{60}{\pi^4} \int\limits_{\risco}^{r_\textrm{out}}\!\!   \frac{r \,\rmd r}{\sqrt{ {\cal D} } }  
 \int_0^{2\pi} \rmd \varphi  \,
\ {\cal L}\ \cos\zeta \ \frac{g^4}{f_{\rm c}^4} \nonumber \\
& \times &
\frac{x'^4 a_{\rm es}(\zeta')}{{\rm e}^{
\displaystyle 
x' t(r,a)/f_{\rm c}} -1} \begin{bmatrix} 1\\
p_{\rm es}(\zeta') \cos(2\chi) \\
p_{\rm es}(\zeta') \sin(2\chi) \end{bmatrix} ,
\end{eqnarray}
where $t(r,a) = r^{3/4} f^{-1/4}(r,a)$ is related to the disk temperature (Eq.~\ref{eq:tempprofile}). 
In this notation, the spectral shape and normalization of $xl_{x}$ are independent of the BH mass and accretion rate.

\begin{figure*}
\centering
\includegraphics[width=0.95\textwidth]{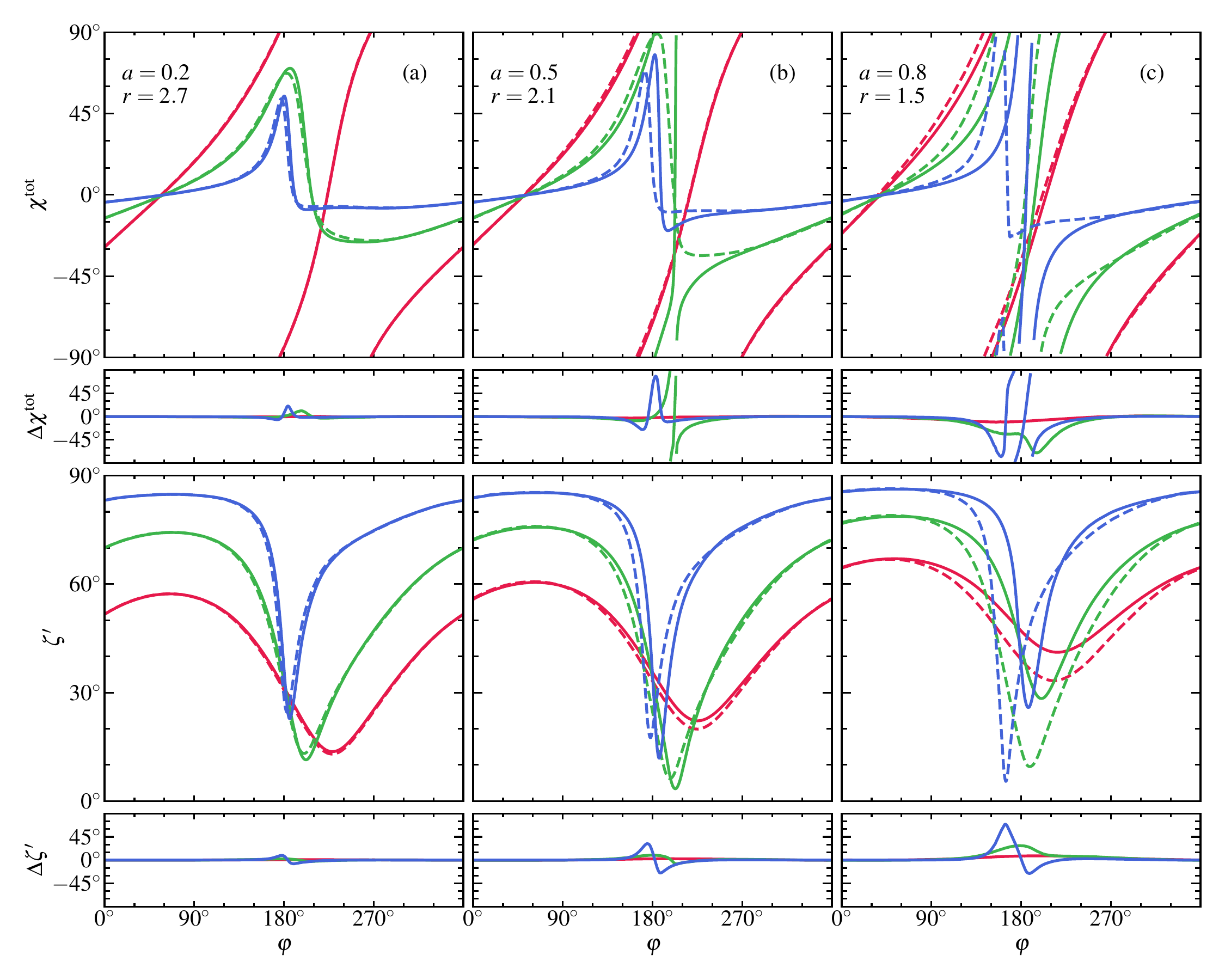}
\caption{Influence of relativistic effects on the PA and the emission angle. 
The rotation of the PA (a sum of the GR and SR effects) $\chi^{\rm tot}$ (\textit{upper panels}) and the emission angle $\zeta'$ (\textit{lower panels}) computed at $r = \risco$ (given in the upper left corner of each panel) for spins $a = 0.2$ (\textit{left panels}), 0.5 (\textit{middle panels}), and 0.8 (\textit{right panels}) and inclinations $i =30\degr$ (red), $60\degr$ (green), and  $80\degr$ (blue lines). 
The results using \textsc{artpol} code are shown with the solid lines, while the results from the ray-tracing in Kerr metric are presented with the dashed lines. 
The difference between the two values is shown under each panel. }
\label{fig:riscodiff}
\end{figure*}

Image of the disk can be reproduced using Cartesian coordinates $X$ and $Y$ (expressed in units of $\rs$) on the plane of the sky:  
\begin{equation}\label{eq:artimage}
\begin{bmatrix}
 X \\ Y   
 \end{bmatrix}
 = 
 b  \begin{bmatrix}
    - \sin\Phi \\
   \cos\Phi
 \end{bmatrix} 
= \frac{r}{\sqrt{\cal D}} \frac{\sin\alpha}{\sin\psi}
 \begin{bmatrix}
  \sin\varphi \\
   - \cos i \cos\varphi
 \end{bmatrix},
 \end{equation}
where $\Phi$ is the position angle of the point where the photon hits the plane of the sky measured counterclockwise from the projection of the disk axis on the sky and 
\begin{equation}\label{eq:impact} 
b =  \frac{r}{\sqrt{\cal D}} \sin\alpha 
 \end{equation}
is the approximation for the impact parameter, which is exact in the Schwarzschild case \citep{PFC83,B02}.

\subsection{Numerical ray-tracing in Kerr metric}
\label{ssec:rt} 

Let us now compare our results to numerical ray-tracing calculations to determine the accuracy of \textsc{artpol}.
We construct the image of the BH accretion disk in the Kerr metric using the \textsc{arcmancer} code \citep{PMN18}, designed to integrate the exact geodesic equation numerically.
The resulting trajectories are parametrized by the point at which they intersect the (observer) image plane.

\begin{figure*}
\centering
\includegraphics[width=0.85\textwidth]{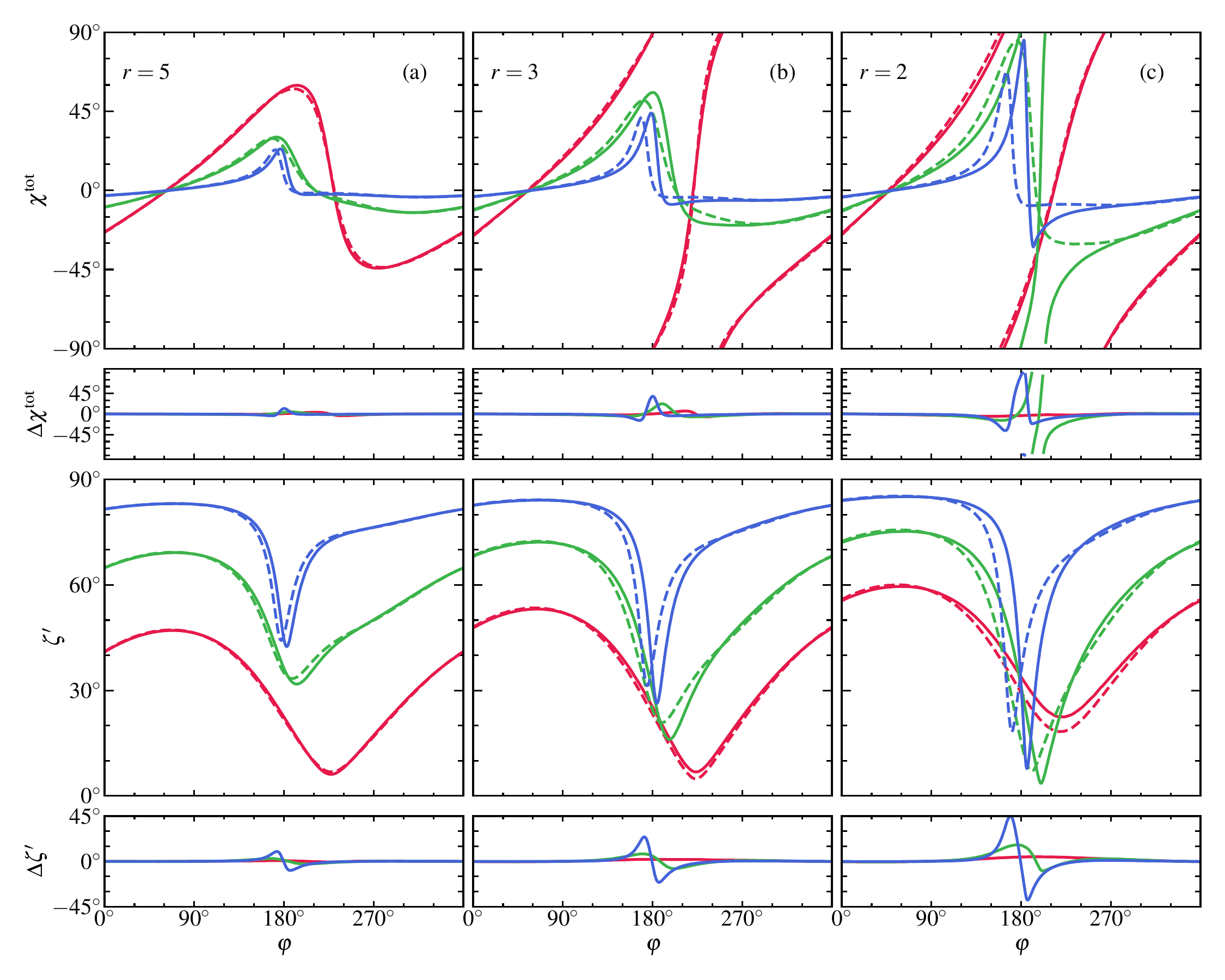}
\caption{Same as Fig.~\ref{fig:riscodiff} for the spin of $a=0.8$ and three ring radii of $r=5$ (\textit{left panels}), 3 (\textit{middle panels}), and 2 (\textit{right panels}).}
\label{fig:rdiff}
\end{figure*}

To obtain an image of the accretion disk, we first define the region of interest on the image plane to include the relevant region of the disk. 
The image plane is positioned at a distance $r_{\rm img} = 2500$, where the effects of general relativity on the photon trajectory are negligible.
We then introduce a grid of nodes and calculate the observed radiation intensity from the accretion disk at each node. 
The rectangular grid on the image plane is orthogonal to the vector pointing toward the observer.
At each point on the grid, we define a local Lorentz frame with the basis vectors along Boyer-Lindquist coordinate grid.
The Lorentz frames constitute the initial conditions for the geodesic curves with the tangent vectors normal to the image plane.
Essentially, a set of parallel rays are propagated from the points of the grid towards the BH, backward in time, until they either intersect the disk surface or reach a distance $>\!r_{\mathrm{img}}$ (as measured from the BH). 
When all the geodesics are calculated, we compute the local intensity and polarization for each point that intersects the disk surface, which we assume lies in the equatorial plane of the Kerr BH.
We consider the velocity and energy dissipation profile following \citet{NT73}.

The Stokes parameters are parallel transported in the ``polarization frame'' along the geodesic in Kerr space-time.
The polarization frame $\mathcal{P}=\{{v}^a,{h}^a\}$ is a pair of orthogonal space-like vectors that are orthogonal to the geodesic itself and the four-velocity of the observer.
Initially, at the image plane, the polarization frame consists of the corresponding two vectors of the local Lorentz frame.
Then, the polarization frame is parallel-transferred along the geodesic to the disk surface to obtain the polarization parameters. 
Then, the frame is projected to the rest frame of the fluid in the disk (comoving frame).
The propagation of the geodesics, transporting of the Lorentz frames, and the projection of the screen to the comoving frame are automatically performed by \textsc{arcmancer}.

All the necessary values, namely, the redshift $g$, the rotation of the polarization basis $\chi^{\rm tot}$, and the emission angle $\zeta'$ can be defined in the comoving frame, where the geodesic hits the disk surface. 
The gravitational redshift factor $g$ is computed using the curve tangents of the initial and the endpoints of the light trajectory.
For more details on this numerical procedure, we refer to Sect.~4 of \citet{PMN18}.
We also correct the redshift by the factor $g_0 = \sqrt{1-1/r_\text{img}}$ to mitigate the redshift between the observer at $r_\text{img}$ and the one at infinity. 
The total redshift agrees with the one given by the analytical expression given by Eq.~\eqref{eq:redshift}.

To compute the overall flux from the disk in the case of numerical ray-tracing, we sum the intensities over the whole image grid as 
\begin{equation} \label{eq:polfluxnum}
 \bm{F}_E = \frac{\rs^2}{D^2}  
 \iint\limits_{r<r_\textrm{out}}     \rmd x \ \rmd y\ 
 {\bf M}(x,y)\ \bm{I}'_{E'}(r, \zeta') ,
\end{equation}
where $ \rmd x\,\rmd y $ is the size of the pixel at $r_\text{img}$, measured in Schwarzschild radii, ${\bf M}(x,y)$ is the rotation (Mueller) matrix  along the trajectory from the point $(x,y)$ on the image plane to the point on the disk surface with a coordinate $r<r_\textrm{out}$.

The secondary, and higher-order images of the disk and its bottom side are visible through the gap between the $\risco$ and the BH event horizon.
From there, a segment of the disk surface can be seen multiple times, depending on how many revolutions its photons make around the BH before escaping toward the observer.
These trajectories are not accounted for in \textsc{artpol} because we find the contribution of the secondary images to the observed flux are below 0.5\% (see Sect.~\ref{sec:spectra} below) for any inclination and spin values \citep[see also][]{Zhou2020}.
Lastly, we note that it is also possible to compute polarization rotation using Walker–Penrose theorem (e.g., see the appendix in \citealt{LiNarayan2009}); however, it does not significantly decrease the computing time 
as long as the geodesics are computed numerically anyway.

\begin{figure*}
\centering
\includegraphics[width=1.0\textwidth]{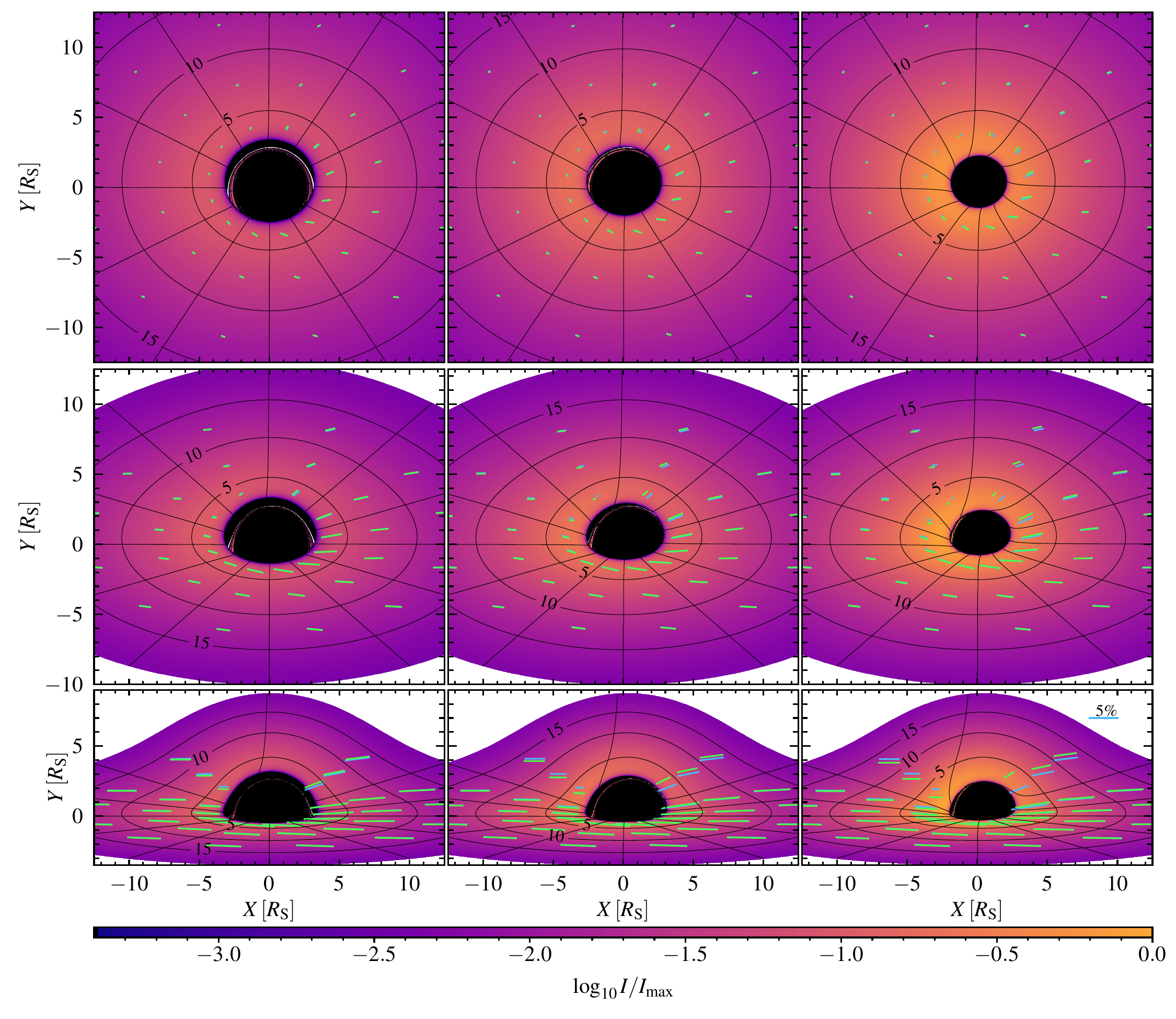}
\caption{Polarization parameters of radiation from the disk at inclinations $i=30\degr$ (\textit{upper row}), $60\degr$ (\textit{middle row}), and  $80\degr$ (\textit{lower row}), and different spin parameters $a=0.2$ (\textit{left column}), 0.5 (\textit{middle column}), and 0.8 (\textit{right column}).
The black lines outline an even polar grid on the disk with rays spaced by $30\degr$ in azimuth and contours spaced by $5 \rs$ in radius (the corresponding values are denoted in each ring).
The sticks represent the polarization parameters derived for the center of each grid segment. 
The polarized disk emission were calculated using the exact numerical interaction of the geodesics in the Kerr metric (with the \textsc{arcmancer} code; green lines) and using the approximate analytical formulae in the Schwarzschild metric (with the \textsc{artpol} code; cyan lines).
The coordinates of the cyan sticks were computed using Eq.~\eqref{eq:artimage}.
The length of the sticks is proportional to the observed PD from the region, and orientation shows the observed PA. 
The 5\% polarization stick is shown in the bottom right panel to scale.  
The colors reflect the logarithm of the bolometric intensity relative to the maximal value across all panels.  
Note that differences between the exact numerical solution and the approximate \textsc{artpol} method are visible only at the inner radii for systems with the highest spin values.}
\label{fig:mapdiff}
\end{figure*}

\section{Applications}
\label{sec:results}

In this section, we show the applications of the \textsc{artpol} technique to spectra and polarization signatures of matter near a Kerr BH.
We first verify the accuracy of the \textsc{artpol} for the PA rotation $\chi$ and the emission zenith angle $\zeta'$ (measured in the comoving frame) for the case of the narrow disk ring located at different distances from the BH.
We proceed to the comparison of polarized images of the accretion disk.
Finally, we show the spectro-polarimetric energy distributions of the accretion disk obtained with analytical and numerical approaches.
In all cases, we assume that the matter (ring/disk) rotates counterclockwise.
Positive spin values correspond to the prograde rotation (BH spin vector aligned with the orbital momentum of the accretion disk).

\begin{figure*}
\centering
\includegraphics[width=1\textwidth]{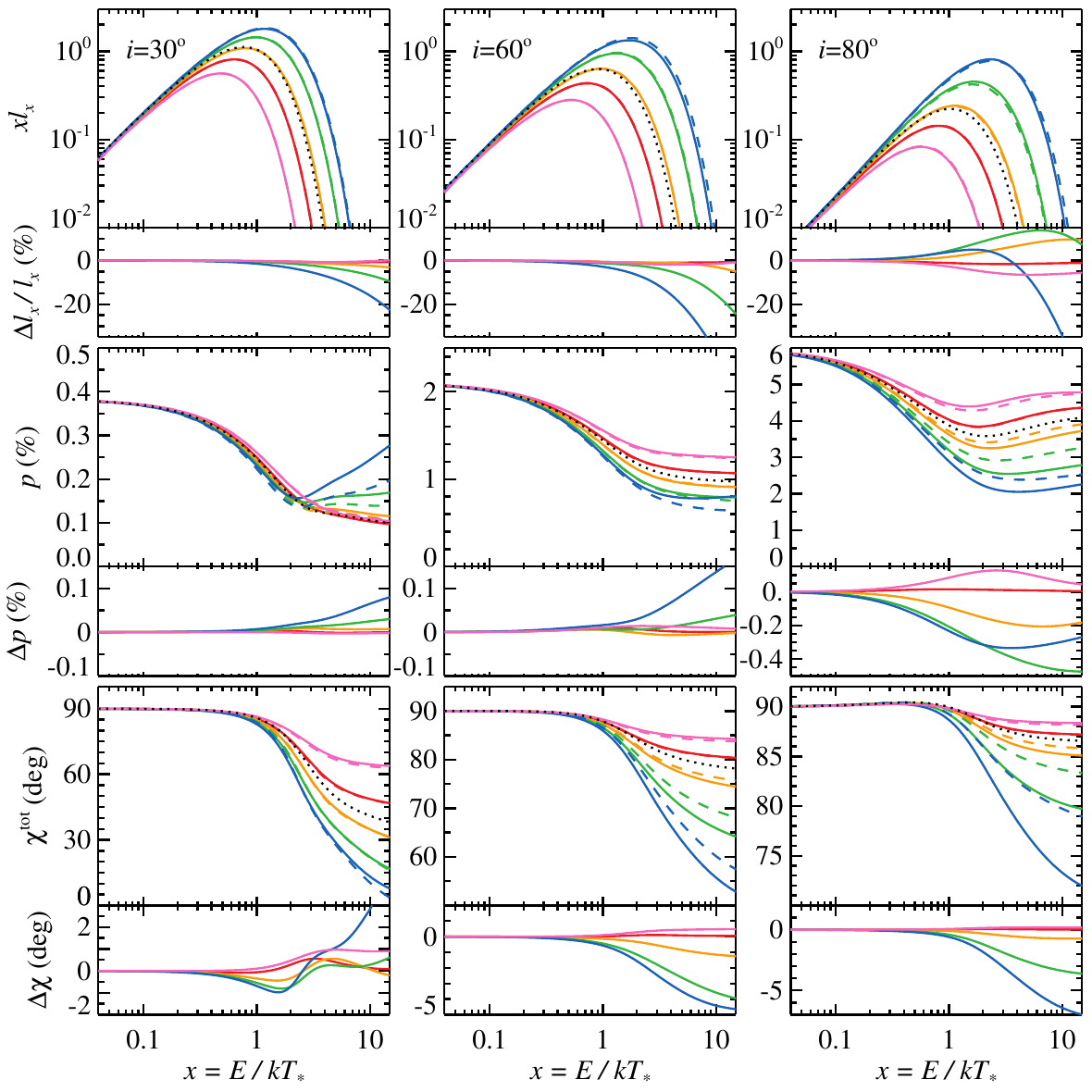}
\caption{Spectra of the luminosity $xl_x$ (upper panels), PD (middle panels) and PA (bottom panels) for different inclinations $i=30\degr$ (left), $60\degr$ (middle), and  $80\degr$ (right column) and BH spin $a=-1$ (pink lines), $0$ (red), 
$0.5$ (orange), $0.8$ (green), and $0.94$ (blue). 
The solid lines correspond to the results obtained with the approximate analytical formulae (\textsc{artpol}), while the dashed lines correspond to the exact numerical integration of the geodesics (\textsc{arcmancer}). 
The accretion disk extends from $\risco$ to  $r_{\rm{out}} = 3000$ for each case. 
Narrow panels show the difference between the results given by the two methods.
The black dotted lines denote the results for a Newtonian disk from \citet{Loktev2022}.}
\label{fig:papdspecs}
\end{figure*}

\subsection{Narrow ring}

The difference between Stokes parameters computed via \textsc{artpol} and numerical ray tracing is expected to be high at the $\risco$, as the frame-dragging effects -- omitted in \textsc{artpol} -- are most important here.
In Fig.~\ref{fig:riscodiff}, we show the total rotation angle $\chi^{\rm tot}$ and the emission zenith angle $\zeta'$ at the ISCO computed by two methods for different spin values and observer inclinations.
For \textsc{artpol}, both PA rotation and $\zeta'$ are systematically shifted, over the azimuth, with respect to the numerical values. 
The difference is highest at $\varphi\sim180\degr$ and, in general, is larger for higher inclinations.
The effect is caused by the close BH approach of the photon trajectories starting from regions of the disk located behind the BH.
Naturally, the difference is larger for higher spin values.
For example, in the case of $a= 0.8$ ($\risco \approx 1.45$) and $i = 30\degr$ the maximum $|\Delta\chi^{\rm tot}|$ is just $10\degr$ and $|\Delta\zeta'|$ is only $9\degr$. 
We also note here that the largest error on $\chi^{\rm tot}$ corresponds to a small emission angle $\cos\zeta'\sim 1$ at which PD is minimal (see Eq.~\ref{eq:Chand_PD}); therefore, the polarized flux is affected very little.

While the accuracy of \textsc{artpol} can be low for certain parts of the narrow ring of matter at the ISCO, for the case of the zero torque boundary condition \citep{NT73}, the total flux emerging from this ring is zero.
The error is then weighted with the dissipation profile, which achieves maximum at distances further than ISCO.
In Fig.~\ref{fig:rdiff} we show $\chi^{\rm tot}$ and $\zeta'$ for $a=0.8$ at different radii $r = 2$, 3, and 5, computed with \textsc{artpol} and numerical ray-tracing. 
The shapes of the lines are similar, but the deviations between the two methods are much smaller. 
For example, the maximum $|\Delta\chi^{\rm tot}|$ and $|\Delta\zeta'|$ for $r=2$ and $i=30\degr$ are $5\degr$ and $4\degr$, respectively, and those are peak values only at $\varphi \approx 180 \degr$.
We note that for $r>5$, \textsc{artpol} gives $\chi^{\rm tot}$ and $\zeta'$ that are nearly indistinguishable from the numerical results, even for BHs with extreme spins.

\subsection{Disk imaging} \label{sec:imaging}

\textsc{artpol} can be used to construct images of the disk (see more details in \citealt{Loktev2022}).
In Fig.~\ref{fig:mapdiff}, we show the observed intensity and polarization of an accretion disk for BH spin parameters $a=0.2$, 0.5, and 0.8, and observer inclinations of $i=30\degr$, 60\degr, and 80\degr. 
The observed intensity for the innermost parts of the standard disk with the electron scattering dominating atmosphere is color-mapped. 
The sticks depict the polarization vector computed with the ray-tracing and analytical techniques in the Schwarzschild metric.
The difference in polarization parameters can barely be seen only in the innermost regions for high spin cases.
Otherwise, the effect of BH rotation is so tiny that the sticks visually coincide.
The sticks are displaced predominantly in the azimuthal direction, mainly caused by the frame-dragging effect, while the radial displacement due to the approximate impact parameter estimation is negligible. 
The most distorted region of the disk appears to be directly behind the black hole, where the surface of the disk appears to be dragged with the BH spin. 
This effect is, again, especially pronounced for higher observer inclinations.

\subsection{Disk spectra and polarization}\label{sec:spectra}

Using the \textsc{artpol} technique, we perform fast computations of the spectral properties, which are essential to extract the spin information in both reflection spectroscopy and continuum-fitting methods and spectral dependence of the polarization signatures.
Comparison of the computed spectral energy distribution, PD, and PA for the case of the optically thick, razor-thin accretion disk to the exact ray-tracing calculations are shown in Fig.~\ref{fig:papdspecs}.
The local model for the disk is described in Sect.~\ref{sec:local_disk}.

In the upper panels of Fig.~\ref{fig:papdspecs}, we show the multicolor blackbody accretion disk spectra viewed at inclinations $i=30\degr$ (left column), $60\degr$ (middle column) and $80\degr$ (right column).
We consider different BH spins from a retrograde one $a=-1$ to the prograde $a=0.94$. 
The solid lines correspond to the calculations using \textsc{artpol}, and the dashed lines correspond to numerical ray-tracing calculations.
The differences between the methods are shown below each panel. 
The deviations increase with the spin value, energy, and inclination, mainly due to the increased importance of the frame-dragging effects.
These are more important for the innermost radii, contributing to higher energies and the photon trajectories passing close to the BH, which appear mostly at higher inclinations.
For higher spin values, the ISCO radii are smaller, and the frame-dragging effect is more pronounced.
At the same time, in the case of the retrograde BH rotation (BH rotates clockwise in our case), $a=-1$, smaller deviations are caused by the more distant location of the ISCO ($\risco=4.5$).
We find that for all the considered spin values, the deviations in the flux values are smaller than 20\% at energies where the flux is relatively high (i.e., where it drops by less than a factor of 10 from the peak value).  

\begin{figure}
\centering\includegraphics[width=0.95\linewidth]{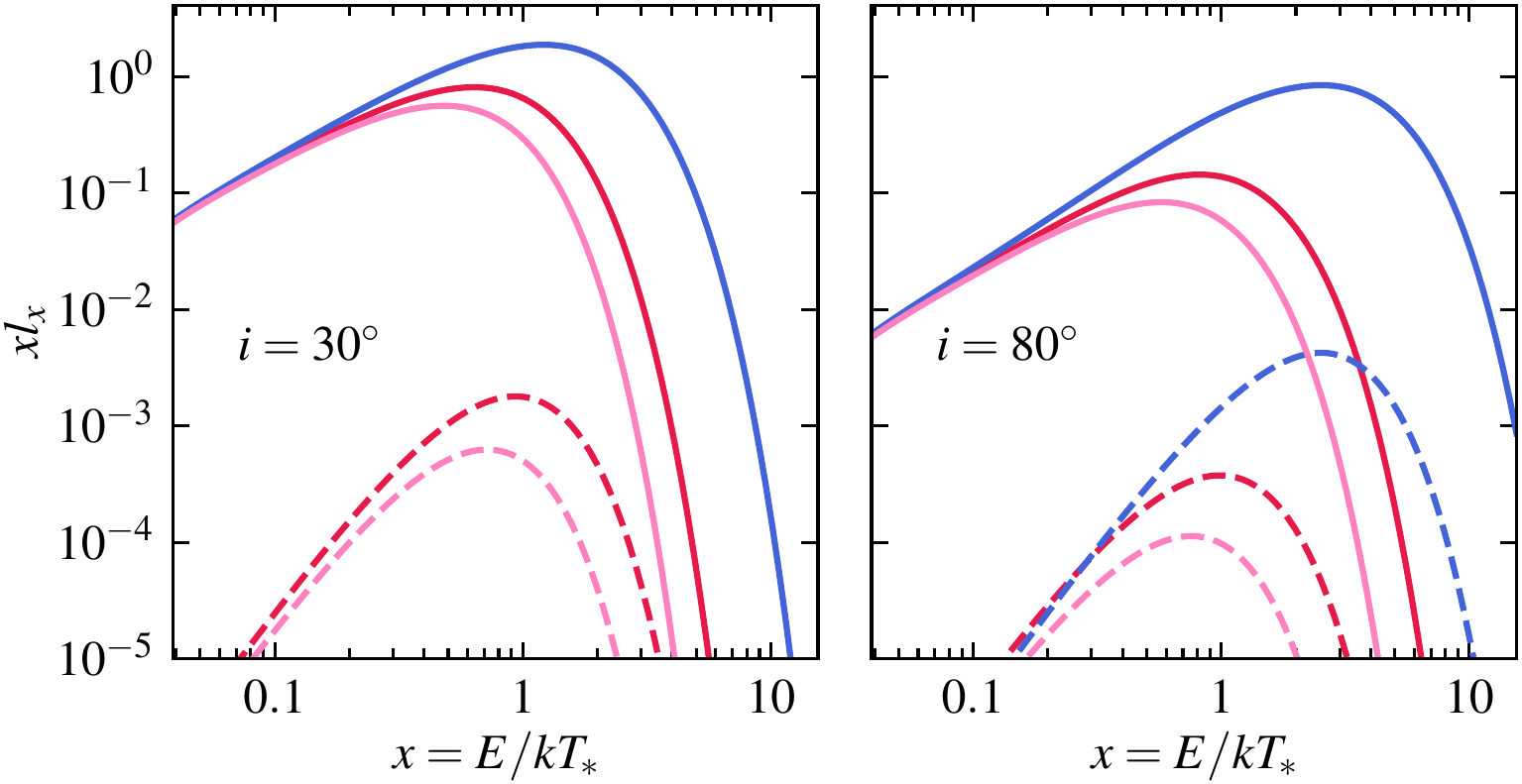}
\caption{Contribution of secondary images to the observed accretion disk flux for two inclinations $i=30\degr$ (\textit{left panel}) and $80\degr$ (\textit{right panel}) and three BH spins $a = -1$ (pink line), $0$ (red), and $0.95$ (blue)  as computed with the numerical ray-tracing code \textsc{arcmancer}. The contribution of the secondary images for $i=30\degr$ and $a=0.95$ is below $10^{-5}$. 
The solid lines correspond to the full luminosity and the dashed lines to the contributions of the secondary images.
}
\label{fig:secondary}
\end{figure}

We can ask here a question whether the difference between the fluxes computed with the two methods has its origin in the presence of secondary images. 
Using \textsc{arcmancer} we computed the accretion disk spectra for two extreme BH spins and two inclinations. 
The results shown in Fig.~\ref{fig:secondary} demonstrate that the secondary images contribute at most 0.5\% of the flux for the high spin BH and large inclination. 
At lower inclinations or smaller spin values, their contribution is even smaller.

The middle and lower panels of Fig.~\ref{fig:papdspecs} show the observed PD and PA. 
The trend of increasing deviations towards higher spins, energies, and inclinations is also evident in these quantities.
For all considered cases, however, the deviations remain small as compared to the typical observational errors achieved in {\it IXPE} observations. 
The maximum deviations in the predicted PD are about 0.4\% for  $a=0.94$ and are much smaller than the expected accuracy of measurements with IXPE for smaller spins.
The error in PA is below 1\degr\ for $a\le 0.5$, and even for $a=0.94$, it is at most 5\degr\ at energies where the flux is relatively high.  
This indicates that fitting the Stokes spectra performed with our analytical technique will work fairly well for all inclinations and all considered spins.

We note the difference between the values obtained for the Newtonian disk \citep{S73} in the Schwarzschild metric in \citet{Loktev2022} and the zero-spin case considered here.
The difference comes solely from the relativistic temperature profile in Eq.~\eqref{eq:tempprofile}, which is different from the Newtonian temperature profile from Eq.~\eqref{eq:corr_Newton} assumed in \citet{Loktev2022}. Notably, the Newtonian disk polarization profiles resemble relativistic profiles with the spin parameter $a \sim 0.2$ more than $a = 0$.

The reduction in calculation time using \textsc{artpol} compared to \textsc{arcmancer} is consistent with that reported in \citet{Loktev2022}, as we are comparing the same two methods. 
Specifically, our calculations of one image using the approximate analytical formulae are performed in $0.01$~s, while the calculations of one polarized spectrum take about $0.05$~s. 
This has to be compared to calculating an image $400\times400$ pixels using a ray-tracing code \textsc{arcmancer}, which takes about $1000$~s.
Moreover, our analytical method offers flexibility, allowing for its application to arbitrary disk geometries, different energy dissipation profiles, and various local spectra and polarization. 

\section{Summary}
\label{sec:summary}

We presented the analytical technique \textsc{artpol} for computing the images and spectro-polarimetric characteristics of relativistic accretion disks around Kerr BHs.
While the velocity and energy dissipation profiles used in the calculations correspond to the space-time around spinning BHs, the photon paths have been considered in the Schwarzschild metric.
We showed that \textsc{artpol} technique is highly efficient, allowing for reduction of the computing time by a factor of $10^4$, and remains accurate, in general within 10\% in flux, 0.2\% in PD, and $1\degr$ in PA, for spin values $a\leq0.5$.
This enables \textsc{artpol} technique to be used for a broad range of BH spin parameters.
The deviations from the exact results obtained by numerical ray-tracing techniques systematically increase towards the highest spins and inclinations, reducing the accuracy down to 20\% in flux, 0.5\% in PD, and $7\degr$ in PA for $a=0.8$--0.94 and $i=80\degr$.
The systematic discrepancy arises from the frame-dragging effect, which is omitted in our analytical ray-tracing calculations.

Applications of the analytical technique include fast computations of static spectro-polarimetric signatures of accretion disks of X-ray binaries in the soft/intermediate/very high states \citep{Ratheesh2023,RodriguezCavero2023,Rawat2023a,Rawat2023b,Jana2023,Podgorny2023}, and account for the relativistic effects on spectra and polarimetric signatures of Comptonization observed in the hard state of X-ray binaries and Seyfert galaxies \citep{Krawczynski22,Marinucci2022,Gianolli2023,Ingram2023,Tagliacozzo2023}.
The long-anticipated detection of the quasi-periodic oscillations from BH X-ray binaries in the polarized light would open new prospects for \textsc{artpol} in terms of fast calculation of the phase-resolved characteristics.
The method is also well suited for the warped disks whose spectro-polarimetric signatures have not been studied in detail so far.

\section*{Acknowledgments}
We acknowledge support from the Jenny and Antti Wihuri foundation (VL), the Academy of Finland grants 355672 (AV), 333112 (JP), and  349144 (VL, AV, JP), and the German Academic Exchange Service (DAAD) project 57525212 (VFS).  
VFS thanks Deutsche  Forschungsgemeinschaft (DFG) grant WE 1312/59-1 for financial support. 

\bibliographystyle{yahapj}
\bibliography{references}

\end{document}